\documentclass[twocolumn,floatfix,prb,showpacs,color,superscriptaddress,epsfig]{revtex4}
\usepackage{graphicx}
\usepackage{epsfig}
\usepackage{amsmath}
\usepackage{epstopdf}
\usepackage{color}

\begin{document}
\title{Engineering topological surface-states: HgS, HgSe and HgTe}

\author{Fran\c{c}ois Virot}
\affiliation{Aix-Marseille Univ., CNRS, IM2NP-UMR 7334, 13397 Marseille Cedex 20, France}
\affiliation{Institut de Radioprotection et de S${\it\hat{u}}$ret\'e Nucl\'eaire,
PSN-RES/SAG/LETR, Centre de Cadarache, 13115 Saint Paul les Durance Cedex, France}
\author{Roland Hayn}
\email{roland.hayn@im2np.fr} \affiliation{Aix-Marseille Univ., CNRS, IM2NP-UMR 7334, 13397 Marseille Cedex 20, France}
\author{Manuel Richter}
\affiliation{IFW Dresden, Helmholtzstr. 20, 01069 Dresden, Germany}
\author{Jeroen van den Brink}
\affiliation{IFW Dresden, Helmholtzstr. 20, 01069 Dresden, Germany}

\date{\today}

\begin{abstract}
Using density functional electronic structure calculations, we establish the consequences of surface termination and modification on protected surface-states of metacinnabar ($\beta$-HgS). Whereas we find that the Dirac cone is isotropic and well-separated from the valence band for the (110) surface, it is highly anisotropic at the pure (001) surface. 
We demonstrate that the anisotropy is modified by surface passivation 
because the topological surface-states include contributions 
from dangling bonds. Such dangling bonds exist on all pure
surfaces within the whole class HgX with X = S, Se, or Te 
and directly affect the properties of the Dirac cone.
Surface modifications also alter the spatial location
(depth and decay length) of the topologically protected edge-states
which renders them essential for the interpretation of photoemission data.
\end{abstract}

\maketitle

An attractive perspective for spintronics was opened up by the discovery\cite{Kane05a,Bernevig06,Fu07a} and manufacturing\cite{Konig07,Hsieh08,Xia09} of topological insulators (TIs). Inside, in its bulk, a TI is insulating but on its boundary -- the material's edges, surfaces, or interfaces -- a TI is conducting. The first TI in which the topological character of its edge-states was established is HgTe, a mercury chalcogenide material of the HgX family  (where X=S, Se, Te) with a zinc-blende crystal structure\cite{Konig07}. The understanding of the precise properties of the topological edge-states in the HgX family of TIs -- and an understanding of how these can be altered and engineered --  is of fundamental importance in the field. 
%
It is of course also
of direct relevance to the development of future spintronics applications, which aims to exploit the spin-momentum locking of TI boundary electrons\cite{Wu06,Zhang09} allowing for instance the generation of edge-spin-currents by circularly polarized light\cite{Raghu10,Hosur11,McIver12}. The precise location and electronic structure of the topological surface-states, in particular their dispersion,
play furthermore
%
an essential role in determining their spin-transport characteristics. It is  for instance well-known that the momentum-spin locking usually renders disordered TIs poor spin conductors: When the electronic spin is tied to its momentum each scattering event randomizes the spin\cite{Burkov10,Schwab11}.  It was recently established that the resulting limit on spin relaxation time $\tau_s$ for edge-state electrons strongly depends on their dispersion 
and momentum space anisotropy\cite{Sacksteder12}.

Here we show that for the HgX class of mercury-based 
II-VI semiconductors in the zinc-blende structure there is a strong 
influence of existing dangling bonds on the properties of topological
surface-states. 
Thus, by surface passivation, it is possible to achieve
a strong modification of the
spatial location of the topological states, their rate of decay into
the bulk, the resulting Dirac cone dispersion
and, in particular, the Dirac cone anisotropy.
Such alterations are possible because the {\it presence} of
topological edge-states is absolutely fixed by the topological
invariants of the bulk electronic structure,
but the shape of their dispersion, their location with respect
to the surface, and their orbital character are not.
On the (001) surface, decoration has an immediate effect
on the Dirac cone dispersion and anisotropy, 
which we will demonstrate in detail for HgX (001) surfaces
where hydrogen passivates the dangling bonds.
A comparison between HgS (001) and (110) surfaces reveals
that location and anisotropy of the topological states 
also strongly depend on the surface type.

Bulk HgTe is an inverted
gapless semiconductor, since the Fermi level is located in the 4-fold degenerate $\Gamma_8$ level.
A tetragonal distortion is needed to open a gap, thus converting HgTe into a 
three-dimensional TI\cite{Brune11}.
HgSe is believed to be similar to HgTe as it has an inverted band structure\cite{Svane11} with $\Gamma_6$ lying below $\Gamma_8$.
In contrast, in $\beta$-HgS the 2-fold
degenerate $\Gamma_7$ level lies above $\Gamma_8$ and 
presumably forms the bottom of the conduction band\cite{Delin02},
which is however still a subject of debate.
Some theoretical investigations
found this material at ambient pressure to be a standard semiconductor with the
$s$-like $\Gamma_6$ state in the conduction band\cite{Moon06,Svane11} while others
confirm that it is an inverted semiconductor\cite{Cardona06,Sakuma11,Virot11}.
It should be noted that the conclusions that we will present here on the surface-states and their anisotropy are independent of  how such debates are settled. We focus on the HgX compounds to illustrate the principle of topological surface-state engineering by the 
controlled chemical and crystallographic modification
of these states which should be applicable to TI in general.

To this end we performed density functional based electronic structure calculations using the all electron, full potential FPLO method\cite{Koepernik99,Opahle99} on HgX slabs with and without surface passivation. 
All calculations were done in the local density approximation\cite{PW92}, where
the presented band structures and projected eigenstates were evaluated in
a four-component relativistic mode, while self-consistency was achieved
in a scalar-relativistic mode.
The valence basis set is comprised of the states
Hg (5$s$, 5$p$, 5$d$, 6$s$, 6$p$, 6$d$, 7$s$);
S  (2$s$, 2$p$, 3$s$, 3$p$, 3$d$, 4$s$, 4$p$);
Se (3$s$, 3$p$, 3$d$, 4$s$, 4$p$, 4$d$, 5$s$, 5$p$);
Te (4$s$, 4$p$, 4$d$, 5$s$, 5$p$, 5$d$, 6$s$, 6$p$); and
H  (1$s$, 2$s$, 2$p$).

\begin{figure}
\begin{center}
\includegraphics[width=0.44\textwidth]{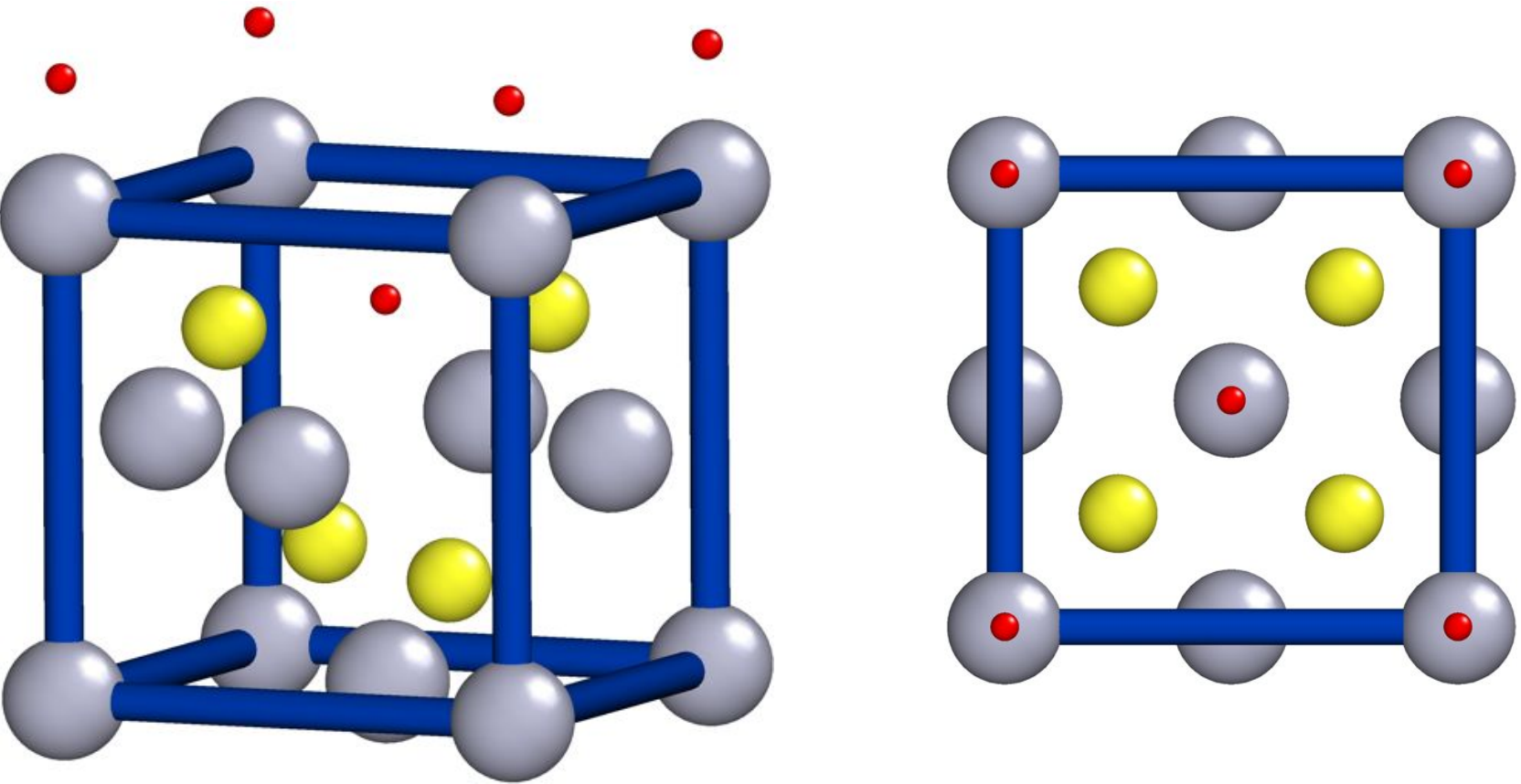}
\end{center}
\caption{(Color online)
Conventional elementary cell of HgX with decorated surface. Left panel: perspective view, right panel: top view onto the (001) surface. Each (HgX)$_4$ cell in the interior of the stack consists of four atomic layers
as shown in the lower part of the cell (Hg: large gray spheres, X: medium yellow
spheres). At the Hg-terminated surface, every second atom is removed. H-atoms (small red spheres) are added in some of the calculations to saturate the dangling bonds.
}
\label{fig:struct}
\end{figure}

In order to model the surfaces, we consider stacks of
(HgX)$_4$ cells, spaced by an empty (vacuum)
layer of $16.5\ldots 25.8$ \AA{} thickness.
Mercury termination is assumed for all (001) surfaces since we found
this termination to be most stable in agreement with the experimental observations on 
HgTe \cite{Oehling98} and HgSe \cite{Eich00}.
To satisfy the stoichiometry, every second Hg-atom has to be removed from the (001) surfaces giving rise to a
$c(2\times2)$ reconstruction, which was observed on HgSe \cite{Eich00} and HgTe \cite{Oehling98}. 
We have considered in addition a decoration of these $c(2\times2)$ (001) surfaces by two H atoms per surface cell, which are placed on top of the surface Hg atoms and in the voids, see Fig.\ \ref{fig:struct}. In these calculations we relaxed the $z$-positions of the decorating H atoms.
For HgS (001), the cells in the stack are cubic with experimental
lattice constant $a = 5.85$ \AA{}.
For (001) stacks of HgSe and HgTe we 
used tetragonally distorted cells, $a \ne c$, with a gap at the $\Gamma$ point.
All (001) slabs have tetragonal symmetry (space group 111).
To build the (110) HgS-slab, 
we used an orthorhombic cell (space group 25 with
stack in $x$-direction) and the same atomic distances
as for the (001) slab. Contrary to (001),
the (110) surface is stoichiometric without any reconstruction. 
The $\bf k$-space integrations were carried out using the linear tetrahedron
method with Bl\"ochl corrections and
$8 \times 8 \times 1$/$4 \times 4 \times 1$ points
in the full Brillouin zone for (001)/(110) slabs.

\begin{figure*}
\begin{center}
\includegraphics[width=0.96\textwidth]{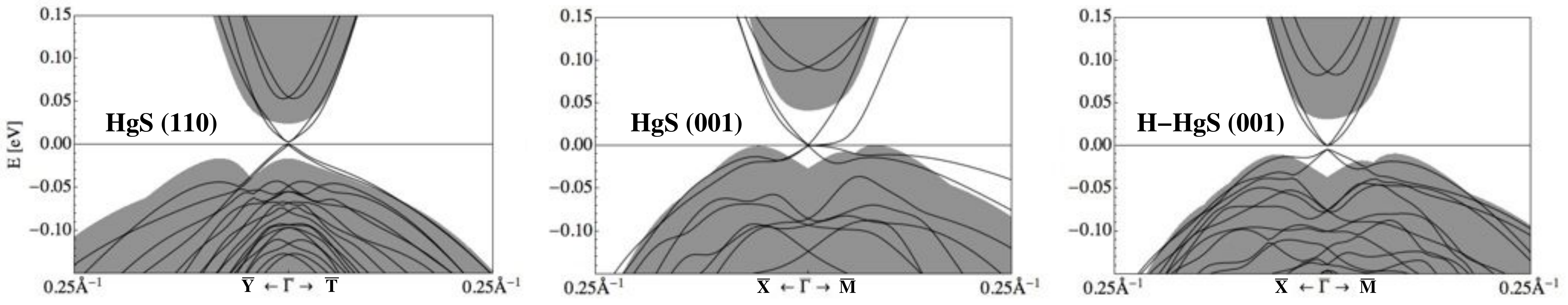}
\end{center}
\caption{Left: Band structure of a 9-cell (110) HgS-slab (lines) close to the
$\Gamma$ point along two specific directions,
superimposed to the HgS bulk band structure (gray area)
projected to the (110) surface;
middle: the same for an 8-cell (001) HgS-slab (same data as published in 
Ref. \onlinecite{Virot11});
right: the same for an 12-cell H-passivated HgS-slab.
}
\label{f1}
\end{figure*}

\begin{figure*}
\begin{center}
\includegraphics[width=0.96\textwidth]{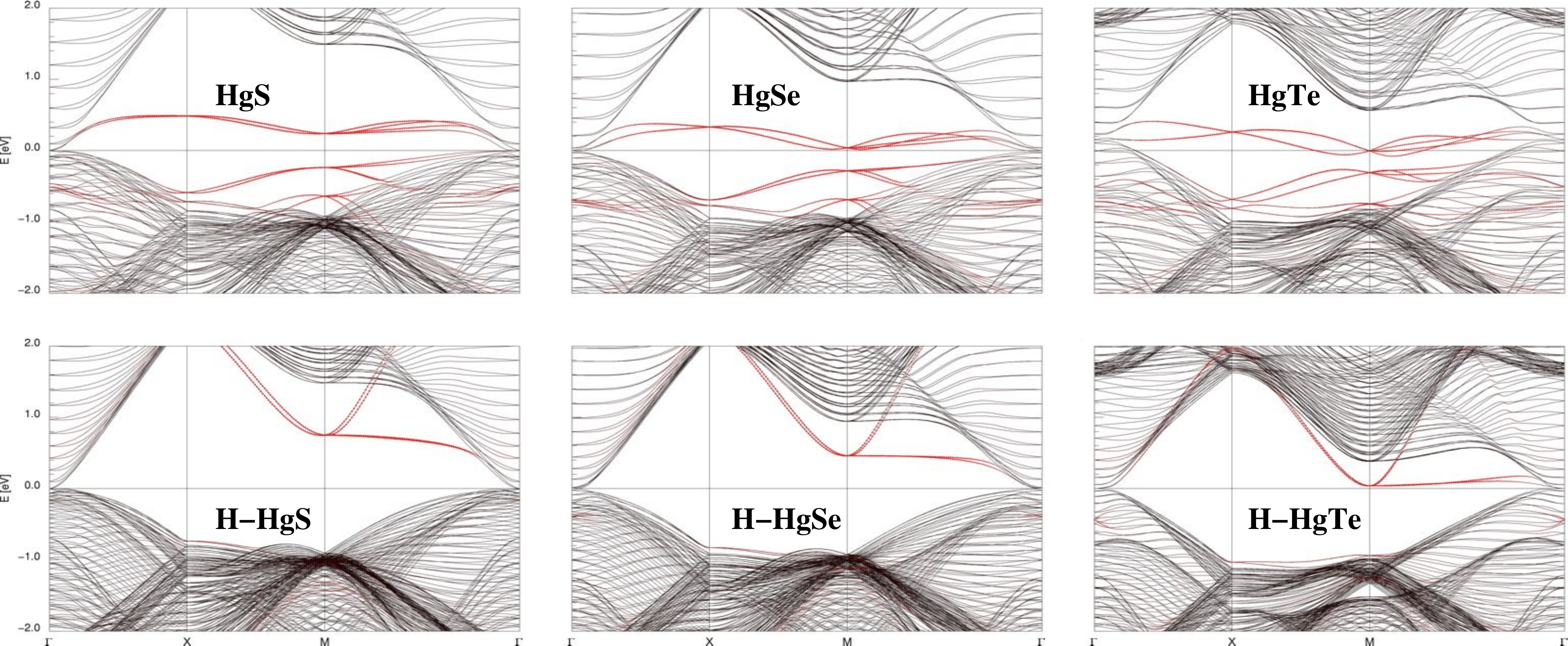}
\end{center}
\caption{(Color online) Band structures of (001) HgX-slabs with zincblende structure.
Upper row: stacks of 8 cells with clean surfaces.
Lower row: stacks of 12 cells with surfaces passivated by hydrogen.
From the left to the right, results for $\beta$-HgS, HgSe, and HgTe are shown. 
The size of red dots on the bands indicates the band weight projected to the 
outermost HgX$_2$(H-HgX$_2$) layers. 
}
\label{f2}
\end{figure*}

We first discuss results of bulk calculations for tetragonally
strained HgSe and HgTe.
Since in cubic HgSe and HgTe there is no gap because only two of the four $\Gamma_8$ levels are occupied,
we consider a strain induced by a CdX (X = Se, Te) substrate.
The in-plane lattice constants equal those of the substrate,
$a = 6.052$ \AA{} (6.48 \AA{}) for X = Se (Te).
In perpendicular direction, $c = 6.13$ \AA{} (6.45 \AA{}) was calculated\cite{Virot12}
using elastic moduli from Ref.\ \onlinecite{Tinjoux03}.
Consequently at the $\Gamma$-point direct gaps open of 24 meV and  10 meV for HgSe and HgTe, respectively.  
This value for HgTe is very close to the measured direct gap of 11 meV\cite{Bouvier11}. There is also a small indirect gap of 5 meV for HgSe at the considered distortion, but HgTe turns out to be metallic in our calculation
in contrast to the recent observation of an indirect gap of 6 meV\cite{Bouvier11}.

Cubic $\beta$-HgS was recently predicted to be a strong TI with 
very anisotropic 
Dirac cones on the pure $c(2\times2)$ (001) surface\cite{Virot11}, 
cf. middle panel of Fig.\ \ref{f1}. 
The anisotropy shows up between the two pairs of occupied/unoccupied
bands at the Fermi level along the direction $\Gamma$-M.
One band of each pair belongs to the top, the other to the bottom
of the slab which has no mirror plane parallel to the surface.
Both bands are interchanged if the $\bf k$-direction is
rotated by $\pi/2$. 

To investigate the influence of the growth direction on the protected surface-states, we calculated the electronic structure of the HgS-(110) surface, which is besides (001) another common surface of the zinc-blende structure\cite{Debias70}. We increased the slab thickness until the residual gap, which is due to the interaction of states at the two opposite surfaces, almost completely closed.
A stack of 9 cells, $74.45$ \AA{} thick, yields a gap of 2.7 meV
at $\Gamma$. 
The corresponding band structure is shown in the left panel of Fig.\ \ref{f1}, where it 
is superimposed onto the projected bulk bands.
As in the case of the (001) surface, 
the projected bulk band structure has a direct gap of 42 meV.
However, there are two remarkable differences between the topological surface-states
of the (110) and the (001) surfaces.
First, the bands cross almost in the center of the bulk gap 
for (110),
while the (001) surface-states cross only slightly 
above the valence band.
Second and most intriguing, the strong anisotropy close to the Dirac point
observed for the (001) surface has almost completely vanished in the (110) case.
This is shown here for the directions $\Gamma$-Y and $\Gamma$-T.
An almost identical dispersion is found along $\Gamma$-Z (not shown).
Note, that there is a mirror plane (glide plane for
even number of atomic layers)
parallel to the surface present in the (110) slab, in contrast to (001). 
Thus, the top/bottom pairs of states are split only
by the residual interaction through the finite slab.
The anisotropy is now visible in a (marginal) difference between 
$\Gamma$-Y and $\Gamma$-T.

It should be noted that in {\em both} cases, (001) and (110),
there is only a twofold rotational symmetry
present at the surface in contrast to the fourfold roto-inversion 
symmetry of the bulk. 
Thus, a certain anisotropy of the Dirac cones is to be
expected also for the (110) surface. 
There, however, each atom is three-fold coordinated with bonds
forming a regular tripod. The related {\em local} three-fold axes
stick out of the surface plane at an angle of $54.7^\circ$.
In contrast, chains of bonds dominate the local symmetry at the
(001) surface, providing a much stronger second-order
ligand field as it is present in the (110) case.

\begin{figure}[!h]
\begin{center}
\includegraphics[width=0.44\textwidth]{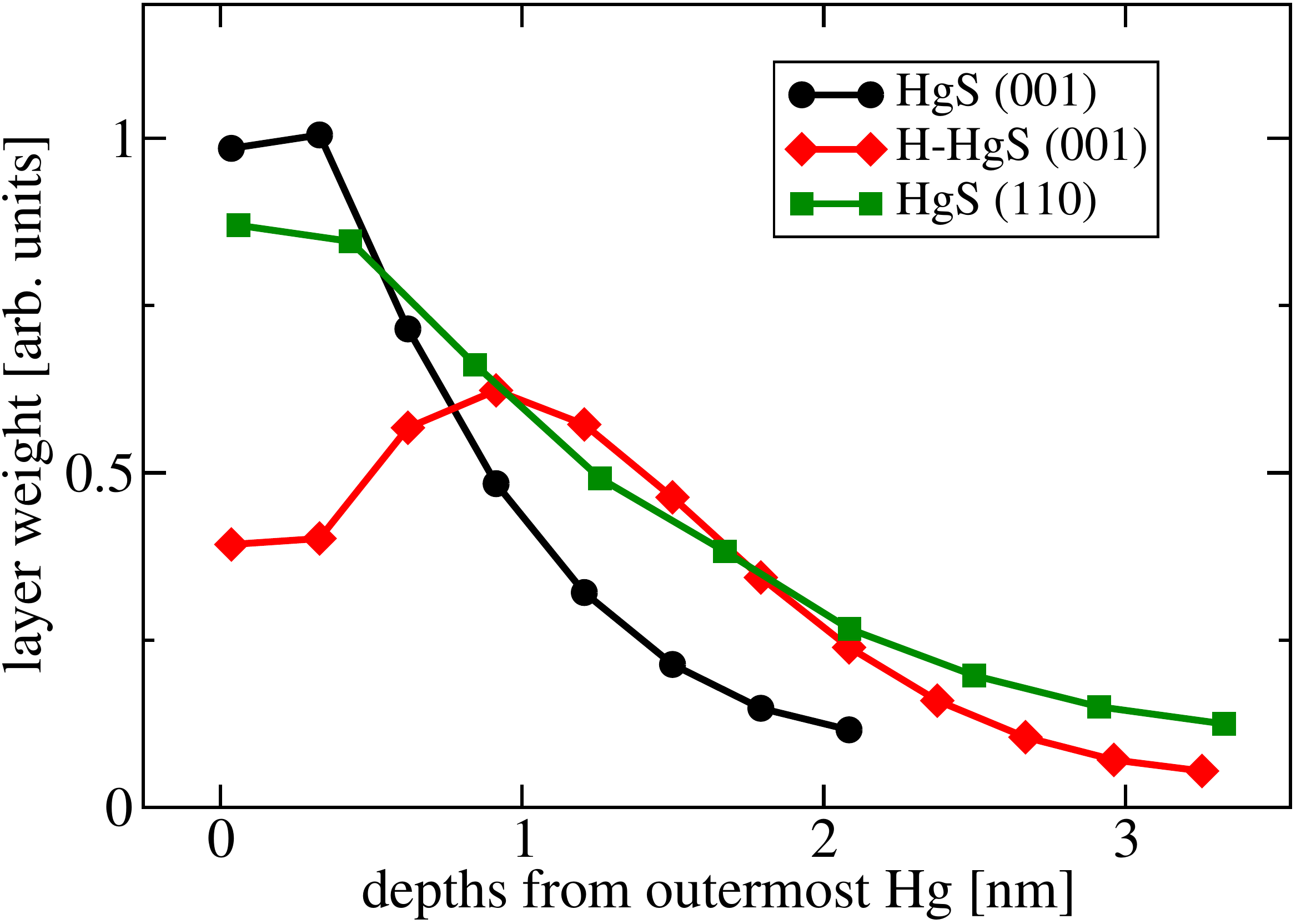}
\end{center}
\caption{(Color online) Layer-resolved weights of the topological surfaces-states
at $\Gamma$. The orbital weights are summed over all four states
(top and bottom of the slab) and over one HgS atomic double-layer
((HgS)$_2$ for (110)).
In the case of H-passivated HgS, the H-weights are included into 
the surface double-layer.}
\label{f5}
\end{figure}

Next, we compare the band structures of (001)
HgX-slabs with clean surfaces, X = S, Se, Te, shown in
the upper row of Fig.\ \ref{f2}.
As a generic feature, eight states with small dispersion are found
in the bulk gap for all three systems.
The upper, unoccupied four states are separated from the bulk continuum
in the whole $\bf k$-space,
while the lower states partly merge with the continuum states. 
By analysis of the atomic contributions (size of red dots in Fig.\ \ref{f2}) we assign these eight states
to dangling bonds which appear as remainders, if
covalent bonds are cut at a surface.
On the (001) surface of HgX, both mercury bonds and bonds
involving $p$-orbitals from the first X sub-surface layer 
are broken.
These dangling bonds of the Hg-terminated (001) surface give rise to dispersionless surface 
bands\cite{Bryant87,Schick89}.
The Dirac cones emerging from the dangling bonds were found to be strongly
anisotropic for HgS\cite{Virot11}, middle panel of Fig.\ \ref{f1}.
HgSe, middle panel of upper row in Fig.\ \ref{f2}, has a very similar
overall electronic structure but no distinct Dirac cone for the present slab thickness 
due to the much smaller bulk gap. 
As mentioned before, HgTe is a bulk metal in our calculation.
Using a ten times larger uniaxial strain which safely opens a finite gap,
Dirac cones were obtained on differently terminated (001) surfaces of HgTe
by applying the {\em ab initio} based method of 
maximally localized Wannier functions in Ref.\ \onlinecite{Yan12}. 
On the Hg-terminated surface, Dirac points with very small velocities were
found at $\bar{\rm X}$ and $\bar{\rm M}$\cite{Yan12}. This finding we confirm in a calculation for the case of realistic strain, see upper right panel of Fig.\ \ref{f2}.

Next, we consider the effect of surface passivation by addition of hydrogen on the 
electronic structure of the (001) surfaces.
Relaxation of the respective $z$-coordinates of H, Fig.\ \ref{fig:struct},
yields a position 0.24 \AA{} (0.27 \AA{}, 0.31 \AA{})
above the subsurface sulfur (Se, Te) layer for the hydrogen in the void and 
a distance of 1.65 \AA{} (1.66 \AA{}, 1.67 \AA{}) between the other hydrogen 
and the mercury atom underneath.
The resulting dispersions are shown in the lower row of Fig.\ \ref{f2}.
Comparison with the band structure of pristine surfaces in the upper row
reveals a considerable reconstruction of the surface-states (red dots).
In particular, the passivation strongly enhances the dispersion of
these states in a bigger part of the $\bf k$-space and shifts them away
from the Fermi level. We attribute this to the covalency introduced
by the H-decoration.

The effect of passivation on the Dirac cones is very clearly illustrated by the case of HgS (001), shown in 
the middle and right panels of Fig.\  \ref{f1} and the left
column of Fig.\ \ref{f2}.
The two bands corresponding to the two surfaces of the slab 
are nearly degenerate after passivation -- 
the anisotropy of the Dirac cone has virtually disappeared.
To understand this effect we calculated the weight contributed by
each HgS-layer to the surface-states which build the Dirac cone,
Fig.\ \ref{f5}. 
For a clean (001) surface,
the topological state is predominantly formed by the dangling
bonds and, thus, located at the outermost atomic
layers with a decay length of about 1 nm.
(Note, however, that the decay length of the dangling-bond states
away from the Dirac point, indicated by large red dots in Fig.\ \ref{f2}, 
is smaller than one atomic double-layer of 0.3 nm.) 
For the passivated surface, however, the maximum weight comes 
from layers 1 nm underneath the surface (Fig.\ \ref{f5}). 
There, the crystal potential is already bulk-like and the
influence of the lower symmetry at the surface is marginal.
This explains the very small anisotropy of the Dirac cone.
We assign the change of weight distribution to the strong
covalency introduced by the passivating hydrogen.
For the (110) surface, where less dangling bonds exists than
at (001), we find an intermediate situation with weight maximum
at the surface but about 2 nm decay length.
A concomitant effect of the inward-shift of the weight, compared
with pristine HgS (001), lies in
a stronger interaction between top and bottom surface bands. 
Therefore, the slab thickness had to be increased
in the two other systems
before a reasonable approximation to a Dirac point was obtained.
We note that the type of information contained in Fig.\ \ref{f5}
is essential for the understanding of photoemission experiments
on TI surface-states, since the electron escape depth is limited.

The electronic structures of passivated HgSe and HgTe are very similar
to that of H-HgS, see the middle and right columns of Fig. \ref{f2}.
In all three cases the passivation removes the states close to the M-point from the Fermi 
level.
Therefore, for the passivated surface of HgSe,
any possible Dirac point must appear at $\Gamma$. 
However, due to the tiny bulk gap 
of only several meV, there is a rather long-range hybridization 
between top and bottom surface-states in our slab
calculation and
the gap at $\Gamma$ closes rather slowly upon increasing the slab thickness.
To obtain a very distinct Dirac cone in HgSe one 
would need slabs much larger than 12 unit cells, 
which is beyond the present possibilities of 
high-precision all-electron methods.
Nevertheless, taking into account the results 
for surface passivation in $\beta$-HgS and the similarity
of the electronic structures for all investigated X, we see that it is highly
likely that the expected Dirac cone will show no large anisotropy
for passivated HgSe (001) either. 
This statement also holds for the case of HgTe in a situation with a  finite bulk band gap which can be 
achieved in a calculation by larger strain\cite{Yan12}.

We have shown in conclusion that surface termination and decoration can play 
a crucial role for topological insulators. 
The topological surface-states can be modified with respect to their dispersion,
their spatial location and, perhaps most importantly, with respect to their anisotropy
by using different surface types or by surface decoration.
We observe for example that surface passivation almost entirely destroys 
the anisotropy of the Dirac cone of the pure (001) surface of $\beta$-HgS. 
The notion, that anisotropic Fermi velocities are beneficial for the spin-transport properties of TIs, 
suggests 
that careful surface preparation is needed to keep these desirable properties intact. An intended generation of
anisotropy, e.g. by surface decoration, is an idea that still remains to be evidenced.

We thank C. Ortix, K. Koepernik and M. Ruck for discussion.
This work was performed using HPC from GENCI-CINES (Grant c2012096873).


\begin{thebibliography}{32}
\expandafter\ifx\csname natexlab\endcsname\relax\def\natexlab#1{#1}\fi
\expandafter\ifx\csname bibnamefont\endcsname\relax
  \def\bibnamefont#1{#1}\fi
\expandafter\ifx\csname bibfnamefont\endcsname\relax
  \def\bibfnamefont#1{#1}\fi
\expandafter\ifx\csname citenamefont\endcsname\relax
  \def\citenamefont#1{#1}\fi
\expandafter\ifx\csname url\endcsname\relax
  \def\url#1{\texttt{#1}}\fi
\expandafter\ifx\csname urlprefix\endcsname\relax\def\urlprefix{URL }\fi
\providecommand{\bibinfo}[2]{#2}
\providecommand{\eprint}[2][]{\url{#2}}

\bibitem[{\citenamefont{Kane and Mele}(2005)}]{Kane05a}
\bibinfo{author}{\bibfnamefont{C.~L.} \bibnamefont{Kane}} \bibnamefont{and}
  \bibinfo{author}{\bibfnamefont{E.~J.} \bibnamefont{Mele}},
  \bibinfo{journal}{Phys. Rev. Lett.} \textbf{\bibinfo{volume}{95}},
  \bibinfo{pages}{146802} (\bibinfo{year}{2005}).

\bibitem[{\citenamefont{Bernevig et~al.}(2006)\citenamefont{Bernevig, Hughes,
  and Zhang}}]{Bernevig06}
\bibinfo{author}{\bibfnamefont{B.~A.} \bibnamefont{Bernevig}},
  \bibinfo{author}{\bibfnamefont{T.~L.} \bibnamefont{Hughes}},
  \bibnamefont{and} \bibinfo{author}{\bibfnamefont{S.-C.} \bibnamefont{Zhang}},
  \bibinfo{journal}{Science} \textbf{\bibinfo{volume}{314}},
  \bibinfo{pages}{1757} (\bibinfo{year}{2006}).

\bibitem[{\citenamefont{Fu et~al.}(2007)\citenamefont{Fu, Kane, and
  Mele}}]{Fu07a}
\bibinfo{author}{\bibfnamefont{L.}~\bibnamefont{Fu}},
  \bibinfo{author}{\bibfnamefont{C.~L.} \bibnamefont{Kane}}, \bibnamefont{and}
  \bibinfo{author}{\bibfnamefont{E.~J.} \bibnamefont{Mele}},
  \bibinfo{journal}{Phys. Rev. Lett.} \textbf{\bibinfo{volume}{98}},
  \bibinfo{pages}{106803} (\bibinfo{year}{2007}).

\bibitem[{\citenamefont{K\"onig et~al.}(2007)\citenamefont{K\"onig, Wiedmann,
  Br\"une, Roth, Buhmann, Molenkamp, Qi, and Zhang}}]{Konig07}
\bibinfo{author}{\bibfnamefont{M.}~\bibnamefont{K\"onig}},
  \bibinfo{author}{\bibfnamefont{S.}~\bibnamefont{Wiedmann}},
  \bibinfo{author}{\bibfnamefont{C.}~\bibnamefont{Br\"une}},
  \bibinfo{author}{\bibfnamefont{A.}~\bibnamefont{Roth}},
  \bibinfo{author}{\bibfnamefont{H.}~\bibnamefont{Buhmann}},
  \bibinfo{author}{\bibfnamefont{L.~W.} \bibnamefont{Molenkamp}},
  \bibinfo{author}{\bibfnamefont{X.-L.} \bibnamefont{Qi}}, \bibnamefont{and}
  \bibinfo{author}{\bibfnamefont{S.-C.} \bibnamefont{Zhang}},
  \bibinfo{journal}{Science} \textbf{\bibinfo{volume}{318}},
  \bibinfo{pages}{766} (\bibinfo{year}{2007}).

\bibitem[{\citenamefont{Hsieh et~al.}(2008)\citenamefont{Hsieh, Qian, Wray,
  Xia, Hor, Cava, and Hasan}}]{Hsieh08}
\bibinfo{author}{\bibfnamefont{D.}~\bibnamefont{Hsieh}},
  \bibinfo{author}{\bibfnamefont{D.}~\bibnamefont{Qian}},
  \bibinfo{author}{\bibfnamefont{L.}~\bibnamefont{Wray}},
  \bibinfo{author}{\bibfnamefont{Y.}~\bibnamefont{Xia}},
  \bibinfo{author}{\bibfnamefont{Y.~S.} \bibnamefont{Hor}},
  \bibinfo{author}{\bibfnamefont{R.~J.} \bibnamefont{Cava}}, \bibnamefont{and}
  \bibinfo{author}{\bibfnamefont{M.~Z.} \bibnamefont{Hasan}},
  \bibinfo{journal}{Nature} \textbf{\bibinfo{volume}{452}},
  \bibinfo{pages}{970} (\bibinfo{year}{2008}).

\bibitem[{\citenamefont{Xia et~al.}(2009)\citenamefont{Xia, Qian, Hsieh, Wray,
  Pal, Lin, Bansil, Grauer, Hor, Cava et~al.}}]{Xia09}
\bibinfo{author}{\bibfnamefont{Y.}~\bibnamefont{Xia}},
  \bibinfo{author}{\bibfnamefont{D.}~\bibnamefont{Qian}},
  \bibinfo{author}{\bibfnamefont{D.}~\bibnamefont{Hsieh}},
  \bibinfo{author}{\bibfnamefont{L.}~\bibnamefont{Wray}},
  \bibinfo{author}{\bibfnamefont{A.}~\bibnamefont{Pal}},
  \bibinfo{author}{\bibfnamefont{H.}~\bibnamefont{Lin}},
  \bibinfo{author}{\bibfnamefont{A.}~\bibnamefont{Bansil}},
  \bibinfo{author}{\bibfnamefont{D.}~\bibnamefont{Grauer}},
  \bibinfo{author}{\bibfnamefont{Y.}~\bibnamefont{Hor}},
  \bibinfo{author}{\bibfnamefont{R.}~\bibnamefont{Cava}}, \bibnamefont{et~al.},
  \bibinfo{journal}{Nature Physics} \textbf{\bibinfo{volume}{5}},
  \bibinfo{pages}{398} (\bibinfo{year}{2009}).

\bibitem[{\citenamefont{Wu et~al.}(2006)\citenamefont{Wu, Bernevig, and
  Zhang}}]{Wu06}
\bibinfo{author}{\bibfnamefont{C.}~\bibnamefont{Wu}},
  \bibinfo{author}{\bibfnamefont{B.~A.} \bibnamefont{Bernevig}},
  \bibnamefont{and} \bibinfo{author}{\bibfnamefont{S.-C.} \bibnamefont{Zhang}},
  \bibinfo{journal}{Phys. Rev. Lett.} \textbf{\bibinfo{volume}{96}},
  \bibinfo{pages}{106401} (\bibinfo{year}{2006}).

\bibitem[{\citenamefont{Zhang et~al.}(2009)\citenamefont{Zhang, Liu, Qi, Dai,
  Fang, and Zhang}}]{Zhang09}
\bibinfo{author}{\bibfnamefont{H.}~\bibnamefont{Zhang}},
  \bibinfo{author}{\bibfnamefont{C.-X.} \bibnamefont{Liu}},
  \bibinfo{author}{\bibfnamefont{X.-L.} \bibnamefont{Qi}},
  \bibinfo{author}{\bibfnamefont{X.}~\bibnamefont{Dai}},
  \bibinfo{author}{\bibfnamefont{Z.}~\bibnamefont{Fang}}, \bibnamefont{and}
  \bibinfo{author}{\bibfnamefont{S.-C.} \bibnamefont{Zhang}},
  \bibinfo{journal}{Nature~Phys.} \textbf{\bibinfo{volume}{5}},
  \bibinfo{pages}{438} (\bibinfo{year}{2009}).

\bibitem[{\citenamefont{Raghu et~al.}(2010)\citenamefont{Raghu, Chung, Qi, and
  Zhang}}]{Raghu10}
\bibinfo{author}{\bibfnamefont{S.}~\bibnamefont{Raghu}},
  \bibinfo{author}{\bibfnamefont{S.~B.} \bibnamefont{Chung}},
  \bibinfo{author}{\bibfnamefont{X.-L.} \bibnamefont{Qi}}, \bibnamefont{and}
  \bibinfo{author}{\bibfnamefont{S.-C.} \bibnamefont{Zhang}},
  \bibinfo{journal}{Phys. Rev. Lett.} \textbf{\bibinfo{volume}{104}},
  \bibinfo{pages}{116401} (\bibinfo{year}{2010}).

\bibitem[{\citenamefont{Hosur}(2011)}]{Hosur11}
\bibinfo{author}{\bibfnamefont{P.}~\bibnamefont{Hosur}},
  \bibinfo{journal}{Phys. Rev. B} \textbf{\bibinfo{volume}{83}},
  \bibinfo{pages}{035309} (\bibinfo{year}{2011}).

\bibitem[{\citenamefont{{McIver} et~al.}(2012)\citenamefont{{McIver}, {Hsieh},
  {Steinberg}, {Jarillo-Herrero}, and {Gedik}}}]{McIver12}
\bibinfo{author}{\bibfnamefont{J.~W.} \bibnamefont{{McIver}}},
  \bibinfo{author}{\bibfnamefont{D.}~\bibnamefont{{Hsieh}}},
  \bibinfo{author}{\bibfnamefont{H.}~\bibnamefont{{Steinberg}}},
  \bibinfo{author}{\bibfnamefont{P.}~\bibnamefont{{Jarillo-Herrero}}},
  \bibnamefont{and} \bibinfo{author}{\bibfnamefont{N.}~\bibnamefont{{Gedik}}},
  \bibinfo{journal}{Nature Nanotechnology} \textbf{\bibinfo{volume}{7}},
  \bibinfo{pages}{96} (\bibinfo{year}{2012}).

\bibitem[{\citenamefont{Burkov and Hawthorn}(2010)}]{Burkov10}
\bibinfo{author}{\bibfnamefont{A.~A.} \bibnamefont{Burkov}} \bibnamefont{and}
  \bibinfo{author}{\bibfnamefont{D.~G.} \bibnamefont{Hawthorn}},
  \bibinfo{journal}{Phys. Rev. Lett.} \textbf{\bibinfo{volume}{105}},
  \bibinfo{pages}{066802} (\bibinfo{year}{2010}).

\bibitem[{\citenamefont{Schwab et~al.}(2011)\citenamefont{Schwab, Raimondi, and
  Gorini}}]{Schwab11}
\bibinfo{author}{\bibfnamefont{P.}~\bibnamefont{Schwab}},
  \bibinfo{author}{\bibfnamefont{R.}~\bibnamefont{Raimondi}}, \bibnamefont{and}
  \bibinfo{author}{\bibfnamefont{C.}~\bibnamefont{Gorini}},
  \bibinfo{journal}{Europhys. Lett.} \textbf{\bibinfo{volume}{93}},
  \bibinfo{pages}{67004} (\bibinfo{year}{2011}).

\bibitem[{\citenamefont{Sacksteder et~al.}(2012)\citenamefont{Sacksteder,
  Kettemann, Wu, Dai, and Fang}}]{Sacksteder12}
\bibinfo{author}{\bibfnamefont{V.~E.} \bibnamefont{Sacksteder}},
  \bibinfo{author}{\bibfnamefont{S.}~\bibnamefont{Kettemann}},
  \bibinfo{author}{\bibfnamefont{Q.~S.}~\bibnamefont{Wu}},
  \bibinfo{author}{\bibfnamefont{X.}~\bibnamefont{Dai}}, \bibnamefont{and}
  \bibinfo{author}{\bibfnamefont{Z.}~\bibnamefont{Fang}},
  \bibinfo{journal}{Phys. Rev. B} \textbf{\bibinfo{volume}{85}},
  \bibinfo{pages}{205303} (\bibinfo{year}{2012}).

\bibitem[{\citenamefont{Br\"une et~al.}(2011)\citenamefont{Br\"une, Liu, Novik,
  Hankiewicz, Buhmann, Chen, Qi, Shen, Zhang, and Molenkamp}}]{Brune11}
\bibinfo{author}{\bibfnamefont{C.}~\bibnamefont{Br\"une}},
  \bibinfo{author}{\bibfnamefont{C.~X.} \bibnamefont{Liu}},
  \bibinfo{author}{\bibfnamefont{E.~G.} \bibnamefont{Novik}},
  \bibinfo{author}{\bibfnamefont{E.~M.} \bibnamefont{Hankiewicz}},
  \bibinfo{author}{\bibfnamefont{H.}~\bibnamefont{Buhmann}},
  \bibinfo{author}{\bibfnamefont{Y.~L.} \bibnamefont{Chen}},
  \bibinfo{author}{\bibfnamefont{X.~L.} \bibnamefont{Qi}},
  \bibinfo{author}{\bibfnamefont{Z.~X.} \bibnamefont{Shen}},
  \bibinfo{author}{\bibfnamefont{S.~C.} \bibnamefont{Zhang}}, \bibnamefont{and}
  \bibinfo{author}{\bibfnamefont{L.~W.} \bibnamefont{Molenkamp}},
  \bibinfo{journal}{Phys. Rev. Lett.} \textbf{\bibinfo{volume}{106}},
  \bibinfo{pages}{126803} (\bibinfo{year}{2011}).

\bibitem[{\citenamefont{Delin}(2002)}]{Delin02}
\bibinfo{author}{\bibfnamefont{A.}~\bibnamefont{Delin}},
  \bibinfo{journal}{Phys. Rev. B} \textbf{\bibinfo{volume}{65}},
  \bibinfo{pages}{153205} (\bibinfo{year}{2002}).

\bibitem[{\citenamefont{Moon and Wei}(2006)}]{Moon06}
\bibinfo{author}{\bibfnamefont{C.-Y.} \bibnamefont{Moon}} \bibnamefont{and}
  \bibinfo{author}{\bibfnamefont{S.-H.} \bibnamefont{Wei}},
  \bibinfo{journal}{Phys. Rev. B} \textbf{\bibinfo{volume}{74}},
  \bibinfo{pages}{045205} (\bibinfo{year}{2006}).

\bibitem[{\citenamefont{Svane et~al.}(2011)\citenamefont{Svane, Christensen,
  Cardona, Chantis, van Schilfgaarde, and Kotani}}]{Svane11}
\bibinfo{author}{\bibfnamefont{A.}~\bibnamefont{Svane}},
  \bibinfo{author}{\bibfnamefont{N.~E.} \bibnamefont{Christensen}},
  \bibinfo{author}{\bibfnamefont{M.}~\bibnamefont{Cardona}},
  \bibinfo{author}{\bibfnamefont{A.~N.} \bibnamefont{Chantis}},
  \bibinfo{author}{\bibfnamefont{M.}~\bibnamefont{van Schilfgaarde}},
  \bibnamefont{and} \bibinfo{author}{\bibfnamefont{T.}~\bibnamefont{Kotani}},
  \bibinfo{journal}{Phys. Rev. B} \textbf{\bibinfo{volume}{84}},
  \bibinfo{pages}{205205} (\bibinfo{year}{2011}).

\bibitem[{\citenamefont{Cardona et~al.}(2006)\citenamefont{Cardona, Kremer,
  Lauck, and Siegle}}]{Cardona06}
\bibinfo{author}{\bibfnamefont{M.}~\bibnamefont{Cardona}},
  \bibinfo{author}{\bibfnamefont{R.~K.} \bibnamefont{Kremer}},
  \bibinfo{author}{\bibfnamefont{R.}~\bibnamefont{Lauck}}, \bibnamefont{and}
  \bibinfo{author}{\bibfnamefont{G.}~\bibnamefont{Siegle}},
 \bibinfo{author}{\bibfnamefont{A.}~\bibnamefont{Munoz}},  
 \bibinfo{author}{\bibfnamefont{A.~H.}~\bibnamefont{Romero}},  
  \bibinfo{journal}{Phys. Rev. B} \textbf{\bibinfo{volume}{80}},
  \bibinfo{pages}{195204} (\bibinfo{year}{2009}).

\bibitem[{\citenamefont{Sakuma et~al.}(2011)\citenamefont{Sakuma, Friedrich,
  Miyaker, Bl\"ugel, and Aryasetiawan}}]{Sakuma11}
\bibinfo{author}{\bibfnamefont{R.}~\bibnamefont{Sakuma}},
  \bibinfo{author}{\bibfnamefont{C.}~\bibnamefont{Friedrich}},
  \bibinfo{author}{\bibfnamefont{T.}~\bibnamefont{Miyake}},
  \bibinfo{author}{\bibfnamefont{S.}~\bibnamefont{Bl\"ugel}}, \bibnamefont{and}
  \bibinfo{author}{\bibfnamefont{F.}~\bibnamefont{Aryasetiawan}},
  \bibinfo{journal}{Phys. Rev. B} \textbf{\bibinfo{volume}{84}},
  \bibinfo{pages}{085144} (\bibinfo{year}{2011}).

\bibitem[{\citenamefont{Virot et~al.}(2011)\citenamefont{Virot, Hayn, Richter,
  and van~den Brink}}]{Virot11}
\bibinfo{author}{\bibfnamefont{F.}~\bibnamefont{Virot}},
  \bibinfo{author}{\bibfnamefont{R.}~\bibnamefont{Hayn}},
  \bibinfo{author}{\bibfnamefont{M.}~\bibnamefont{Richter}}, \bibnamefont{and}
  \bibinfo{author}{\bibfnamefont{J.}~\bibnamefont{van~den Brink}},
  \bibinfo{journal}{Phys. Rev. Lett.} \textbf{\bibinfo{volume}{106}},
  \bibinfo{pages}{236806} (\bibinfo{year}{2011}).

\bibitem[{\citenamefont{Koepernik and Eschrig}(1999)}]{Koepernik99}
\bibinfo{author}{\bibfnamefont{K.}~\bibnamefont{Koepernik}} \bibnamefont{and}
  \bibinfo{author}{\bibfnamefont{H.}~\bibnamefont{Eschrig}},
  \bibinfo{journal}{Phys. Rev. B} \textbf{\bibinfo{volume}{59}},
  \bibinfo{pages}{1743} (\bibinfo{year}{1999}),
  \urlprefix\url{http://www.fplo.de/}.

\bibitem[{\citenamefont{Opahle et~al.}(1999)\citenamefont{Opahle, Koepernik,
  and Eschrig}}]{Opahle99}
\bibinfo{author}{\bibfnamefont{I.}~\bibnamefont{Opahle}},
  \bibinfo{author}{\bibfnamefont{K.}~\bibnamefont{Koepernik}},
  \bibnamefont{and} \bibinfo{author}{\bibfnamefont{H.}~\bibnamefont{Eschrig}},
  \bibinfo{journal}{Phys. Rev. B} \textbf{\bibinfo{volume}{60}},
  \bibinfo{pages}{14035} (\bibinfo{year}{1999}).

\bibitem[{\citenamefont{Perdew and Wang}(1992)}]{PW92}
\bibinfo{author}{\bibfnamefont{J.~P.} \bibnamefont{Perdew}} \bibnamefont{and}
  \bibinfo{author}{\bibfnamefont{Y.}~\bibnamefont{Wang}},
  \bibinfo{journal}{Phys. Rev. B} \textbf{\bibinfo{volume}{45}},
  \bibinfo{pages}{13244} (\bibinfo{year}{1992}).

\bibitem[{\citenamefont{Oehling et~al.}(1998)\citenamefont{Oehling, Ehinger,
  Gerhard, Becker, Landwehr, Schneider, Eich, Neureiter, Fink, Sokolowski
  et~al.}}]{Oehling98}
\bibinfo{author}{\bibfnamefont{S.}~\bibnamefont{Oehling}},
  \bibinfo{author}{\bibfnamefont{M.}~\bibnamefont{Ehinger}},
  \bibinfo{author}{\bibfnamefont{T.}~\bibnamefont{Gerhard}},
  \bibinfo{author}{\bibfnamefont{C.~R.} \bibnamefont{Becker}},
  \bibinfo{author}{\bibfnamefont{G.}~\bibnamefont{Landwehr}},
  \bibinfo{author}{\bibfnamefont{M.}~\bibnamefont{Schneider}},
  \bibinfo{author}{\bibfnamefont{D.}~\bibnamefont{Eich}},
  \bibinfo{author}{\bibfnamefont{H.}~\bibnamefont{Neureiter}},
  \bibinfo{author}{\bibfnamefont{R.}~\bibnamefont{Fink}},
  \bibinfo{author}{\bibfnamefont{M.}~\bibnamefont{Sokolowski}},
  \bibnamefont{et~al.}, \bibinfo{journal}{Appl. Phys. Letters}
  \textbf{\bibinfo{volume}{73}}, \bibinfo{pages}{3205} (\bibinfo{year}{1998}).

\bibitem[{\citenamefont{Eich et~al.}(2000)\citenamefont{Eich, Hubner, Ortner, 
Kilian, Becker, Landwehr, Fink, and Umbach}}]{Eich00}
\bibinfo{author}{\bibfnamefont{D.}~\bibnamefont{Eich}},
  \bibinfo{author}{\bibfnamefont{D.}~\bibnamefont{Hubner}},
  \bibinfo{author}{\bibfnamefont{K.}~\bibnamefont{Ortner}},
  \bibinfo{author}{\bibfnamefont{L.} \bibnamefont{Kilian}},
  \bibinfo{author}{\bibfnamefont{R.}~\bibnamefont{Becker}},
  \bibinfo{author}{\bibfnamefont{G.}~\bibnamefont{Landwehr}},
  \bibinfo{author}{\bibfnamefont{R.}~\bibnamefont{Fink}}, \bibnamefont{and}
  \bibinfo{author}{\bibfnamefont{E.}~\bibnamefont{Umbach}},
  \bibinfo{journal}{Appl. Surface Science}
  \textbf{\bibinfo{volume}{166}}, \bibinfo{pages}{12} (\bibinfo{year}{2000}).


\bibitem[{\citenamefont{Virot}(2012)}]{Virot12}
\bibinfo{author}{\bibfnamefont{F.}~\bibnamefont{Virot}},
  \bibinfo{howpublished}{doctoral thesis, University Aix-Marseille, Marseille
  (France)} (\bibinfo{year}{2012}).

\bibitem[{\citenamefont{Tinjoux}(2003)}]{Tinjoux03}
\bibinfo{author}{\bibfnamefont{F.}~\bibnamefont{Tinjoux}},
  \bibinfo{howpublished}{doctoral thesis, University Joseph Fourier, Grenoble
  (France)} (\bibinfo{year}{2003}).

\bibitem[{\citenamefont{Bouvier et~al.}(2011)\citenamefont{Bouvier, Meunier,
  Kramer, and L\'evy}}]{Bouvier11}
\bibinfo{author}{\bibfnamefont{C.}~\bibnamefont{Bouvier}},
  \bibinfo{author}{\bibfnamefont{T.}~\bibnamefont{Meunier}},
  \bibinfo{author}{\bibfnamefont{R.}~\bibnamefont{Kramer}}, \bibnamefont{and}
  \bibinfo{author}{\bibfnamefont{L.~P.} \bibnamefont{L\'evy}},
  \bibinfo{journal}{ArXiv e-prints}  (\bibinfo{year}{2011}),
  \eprint{1112.2092}.

\bibitem[{\citenamefont{Debias et~al.}(1970)\citenamefont{Debias, Barcelo,
  Aicardi, Masse, and Bombre}}]{Debias70}
\bibinfo{author}{\bibfnamefont{G.}~\bibnamefont{Debias}},
  \bibinfo{author}{\bibfnamefont{J.}~\bibnamefont{Barcelo}},
  \bibinfo{author}{\bibfnamefont{J.~P.} \bibnamefont{Aicardi}},
  \bibinfo{author}{\bibfnamefont{G.}~\bibnamefont{Masse}}, \bibnamefont{and}
  \bibinfo{author}{\bibfnamefont{F.}~\bibnamefont{Bombre}},
  \bibinfo{journal}{Thin Solid Films} \textbf{\bibinfo{volume}{7}},
  \bibinfo{pages}{11} (\bibinfo{year}{1970}).

\bibitem[{\citenamefont{Bryant}(1987)}]{Bryant87}
\bibinfo{author}{\bibfnamefont{G.~W.} \bibnamefont{Bryant}},
  \bibinfo{journal}{Phys. Rev. B} \textbf{\bibinfo{volume}{35}},
  \bibinfo{pages}{5547} (\bibinfo{year}{1987}).

\bibitem[{\citenamefont{Schick et~al.}(1989)\citenamefont{Schick, Bose, and
  Chen}}]{Schick89}
\bibinfo{author}{\bibfnamefont{J.~T.} \bibnamefont{Schick}},
  \bibinfo{author}{\bibfnamefont{S.~M.} \bibnamefont{Bose}}, \bibnamefont{and}
  \bibinfo{author}{\bibfnamefont{A.-B.} \bibnamefont{Chen}},
  \bibinfo{journal}{Phys. Rev. B} \textbf{\bibinfo{volume}{40}},
  \bibinfo{pages}{7825} (\bibinfo{year}{1989}).

\bibitem[{\citenamefont{Yan and Zhang}(2012)}]{Yan12}
\bibinfo{author}{\bibfnamefont{B.}~\bibnamefont{Yan}} \bibnamefont{and}
  \bibinfo{author}{\bibfnamefont{S.-C.} \bibnamefont{Zhang}},
  \bibinfo{journal}{Reports on Progress in Physics}
  \textbf{\bibinfo{volume}{75}}, \bibinfo{pages}{096501}
  (\bibinfo{year}{2012}).

\end{thebibliography}

\end{document}